\documentclass[aps,prl,amsmath,amssymb,reprint,superscriptaddress,preprintnumbers,showpacs,intlimits]{revtex4-1}
\usepackage{bm,latexsym,mathrsfs,enumerate,color}
\usepackage[mathcal]{euscript}
\usepackage[breaklinks=true,unicode=true,urlcolor = blue,colorlinks = true,citecolor = blue,linkcolor = blue]{hyperref}
\usepackage{graphicx}
\renewcommand{\vec}[1]{\bm{#1}}
\begin{document}


\title{Curvature effects in vortex chirality switching}

\author{Kostiantyn V. Yershov}
\email{yershov@bitp.kiev.ua}
\affiliation{Bogolyubov Institute for Theoretical Physics, 03680 Kyiv, Ukraine}
\affiliation{National University of ``Kyiv-Mohyla Academy", 04655 Kyiv, Ukraine}

\author{Volodymyr P. Kravchuk}
 \email{vkravchuk@bitp.kiev.ua}
 \affiliation{Bogolyubov Institute for Theoretical Physics, 03680 Kyiv, Ukraine}

\author{Denis D. Sheka}
\email{sheka@univ.net.ua}
\affiliation{Taras Shevchenko National University of Kyiv, 01601 Kyiv, Ukraine}

%

\author{Yuri~Gaididei}
\email{ybg@bitp.kiev.ua}
 \affiliation{Bogolyubov Institute for Theoretical Physics, 03680 Kyiv, Ukraine}

\date{\today}

%
%

\begin{abstract}
A simple mechanism of controllable switching of magnetic vortex chirality is proposed. We consider curvilinear magnetic nanoshells of spherical geometry whose ground state is a vortex magnetization distribution. Chirality of this magnetic vortex can be switched in controllable way by applying a Gaussian pulse of spatially uniform magnetic field along the symmetry axis of the shell. The chirality switching process is explored in detail numerically for various parameters of magnetic pulse: the corresponding switching diagram is build. The role of the curvature is ascertained by studying the switching diagram evolution under the continuous transition from hemispherical shell to the disk shaped sample with the volume and thickness kept constant.
\end{abstract}


%
%

%
%


\maketitle

Magnetically soft ferromagnetic nanoparticle of sub-micron size and symmetrical form typically demonstrates the ground state in form of a vortex magnetization distribution, this is the result of competition of strong short-range exchange and weak long-range dipole-dipole interactions.\cite{Hubert98} Starting from seminal work of E.~Feldkeller and H.~Thomas\cite{Feldtkeller65} magnetic vortices are widely studied for planar magnets. Also recently it was demonstrated both theoretically\cite{Sheka13b} and experimentally\cite{Soares08,*Streubel12,*Streubel12a} that the vortex state can be ground one for hemispherical magnetic nanocaps.

Magnetic vortex has two binary characteristics: \textit{chirality} $C=\pm1$, the counterclockwise ($C=+1$) or clockwise ($C=-1$) direction of magnetization circulation; and \textit{polarity} $p=\pm1$, the up ($p=+1$) or down ($p=-1$) direction of the vortex core magnetization. Each of these quantities can be potentially used for storing of a bit of information in a high-speed magnetic random access memory.\cite{Bohlens08,*Pigeau10,*Yu11a,*Nakano11} In this respect the possibility of controllable manipulating of the chirality and polarity values is crucial one. Though a number of mechanisms of controllable vortex polarity switching are already proposed,\cite{Waeyenberge06,Hertel07,Kravchuk07c,*Curcic08a,*Kammerer11,Caputo07,*Yamada07} the controllable vortex chirality switching is still a challenging problem because it requires a fine asymmetrical tuning the nanoscale system: the asymmetry must be introduced into geometry of the nanomagnet\cite{Jaafar10,*Uhlir13,*Hertel13,*Lua08} or into the spatial distribution of the applied magnetic fields.\cite{Gaididei08} In symmetrical planar systems e.g. magnetic nanodisks, the vortex chirality control requires highly accurate setting of parameters of the of magnetic field pulse\cite{Antos09} or spin-current pulse.\cite{Choi10}

\begin{figure}
\includegraphics[width=0.9\columnwidth]{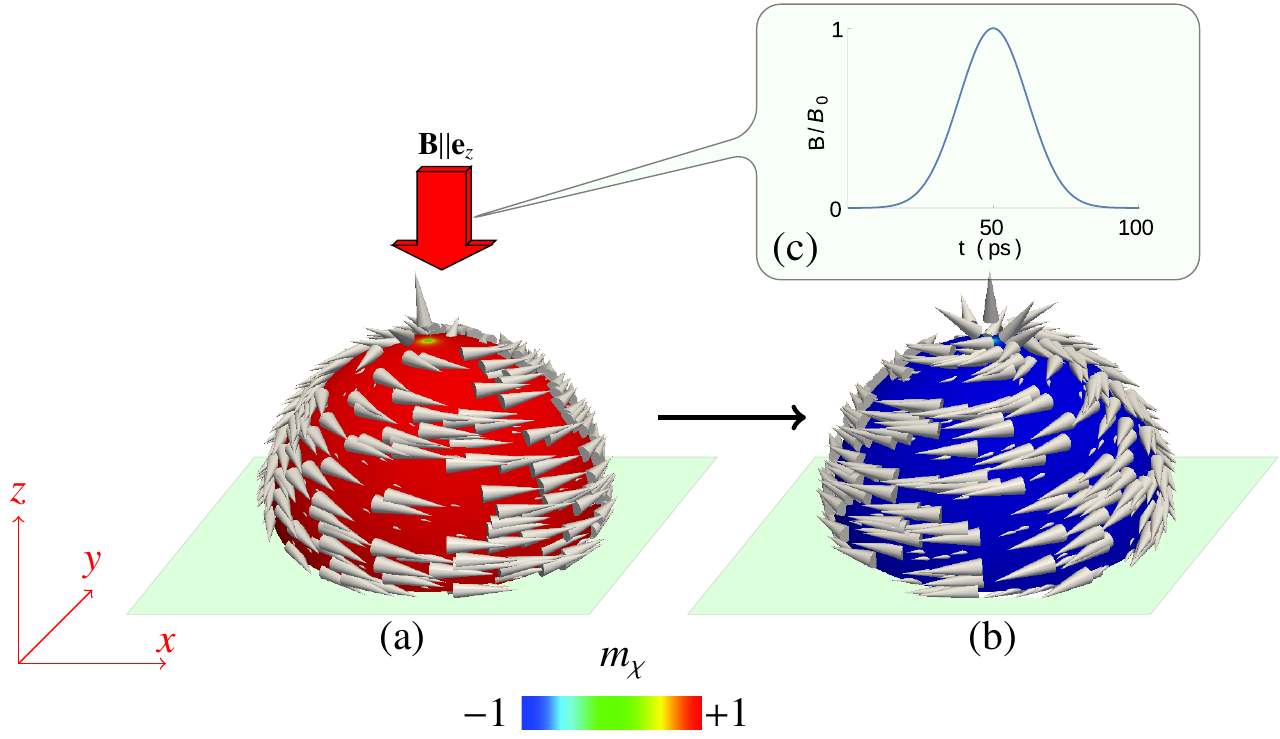}
\caption{(Color online) The scheme of vortex chirality switching for a hemispherical shell. (a) Initial vortex state with $C=1$. (b) Resulting vortex state with $C=-1$. (c) Gaussian temporal profile of a spatially uniform magnetic pulse.}\label{fig:switching-scheme}
\end{figure}

Here we propose a simple mechanism of vortex chirality control for a symmetrical system by using a simple spatially uniform pulse of magnetic field. The idea is to proceed from planar to curvilinear magnetic shells with spherical geometry, see Fig.~\ref{fig:switching-scheme}. We choose the magnetic pulse with the Gaussian temporal profile $\vec B= -\vec e _z B_0\exp[-(t-3\tau)^2/\tau^2]$, where $B_0$ and $\tau$ are amplitude and width of the field pulse respectively. The pulse is applied along the symmetry axis of the system, $\vec B || \vec e_z$. Note, that recently we reported on the vortex polarity switching for spherical shells under action of the same Gaussian field pulse applied within the $xy$-plane\cite{Sloika14}.

\begin{figure*}
\includegraphics[width=\textwidth]{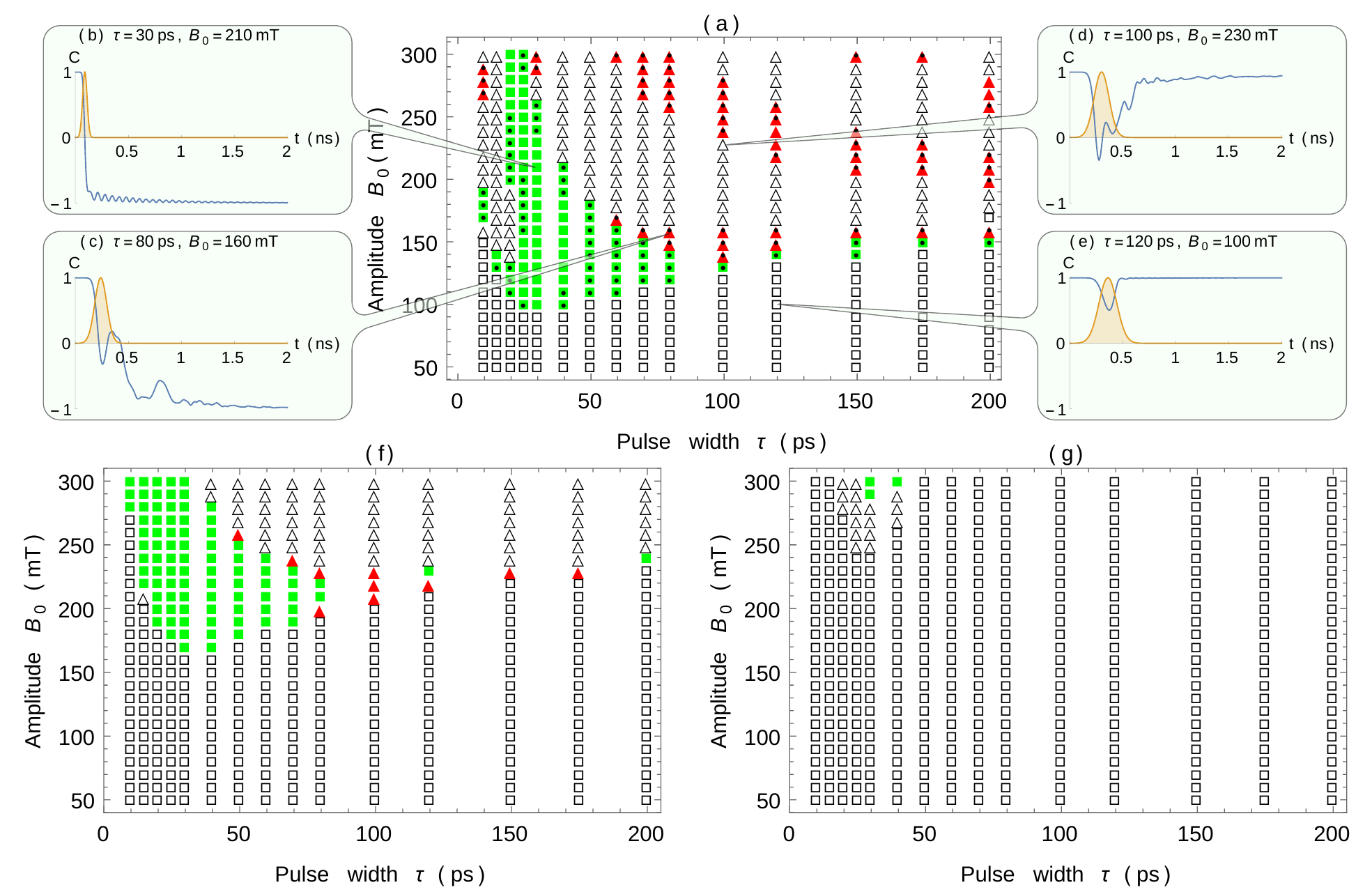}
\caption{(Color online) (a) chirality switching diagram for the hemispherical shell. Filled and open symbols determine the pulse parameters when the switching does and does not take place, respectively. Triangles correspond to potentially uncontrollable switching regimes when $C(t)$ changes its sign more than once. The corresponding chirality time behavior typical for different regimes is demonstrated on insets (b)-(e), the filled area shows the pulse profile. Diagrams (f) and (g) demonstrate evolution of the switching diagram with the curvature radius $R$ increasing, namely (f) corresponds to the cap with $R\approx1.85 R_0$ nm (truncation polar angle $\vartheta_c=\pi/4$) and (e) corresponds to the disk shaped sample with curvature radius $R\to\infty$ and straight radius $R_\mathrm{disk}=\sqrt2 R_0$. Black spots on the inset (a) correspond to the switching mechanism by means of the vortex-antivortex pairs formation, see Fig.~\ref{fig:types_of_switching}(a-f).
}\label{fig:diagrams}
\end{figure*}

\begin{figure}
\includegraphics[width=\columnwidth]{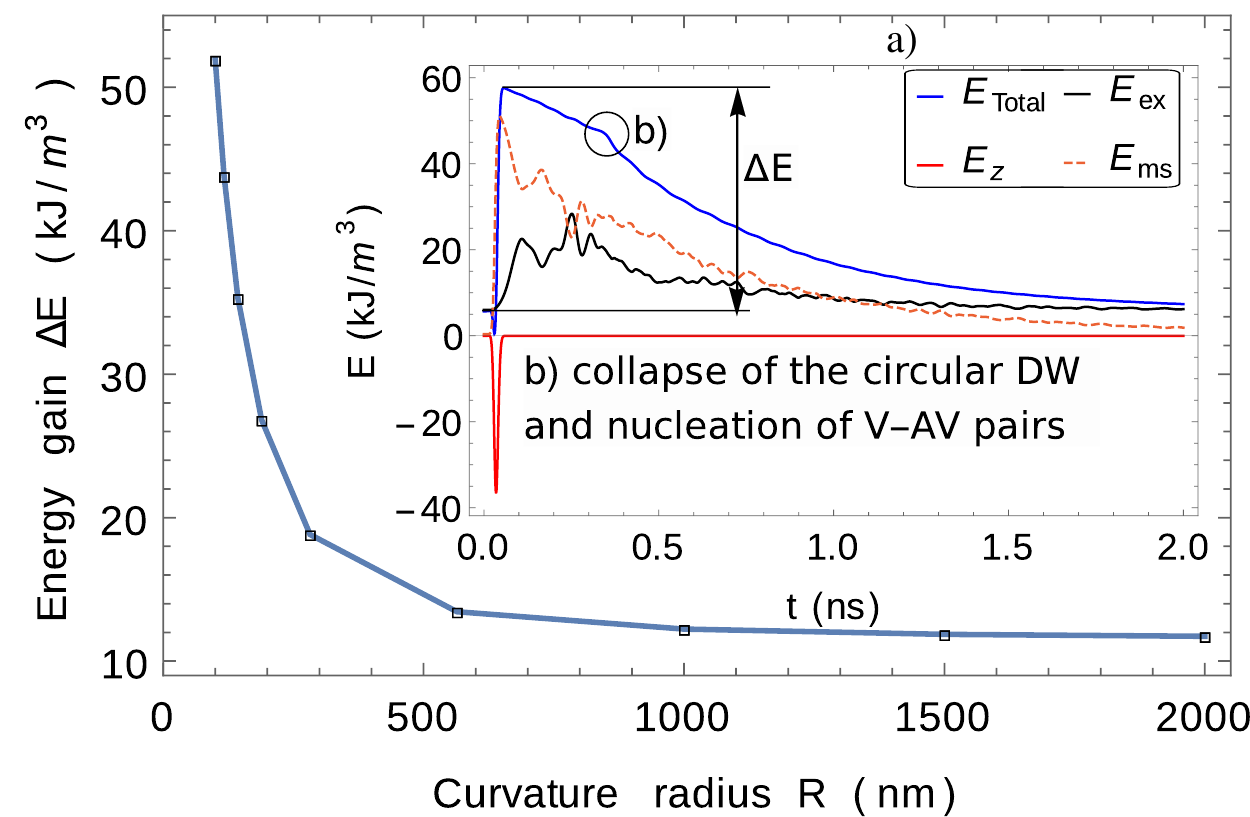}
\caption{(Color online) Dependence of absorbed energy on the curvature radius of the shell under the action of magnetic field pulse with $B_0=180$ mT and $\tau=10$ ps. The inset (a) demonstrates time evolution of different energy constituents for the case $R=100$ nm.
}\label{fig:energies}
\end{figure}

\begin{figure*}[h!bt]
\includegraphics[width=\textwidth]{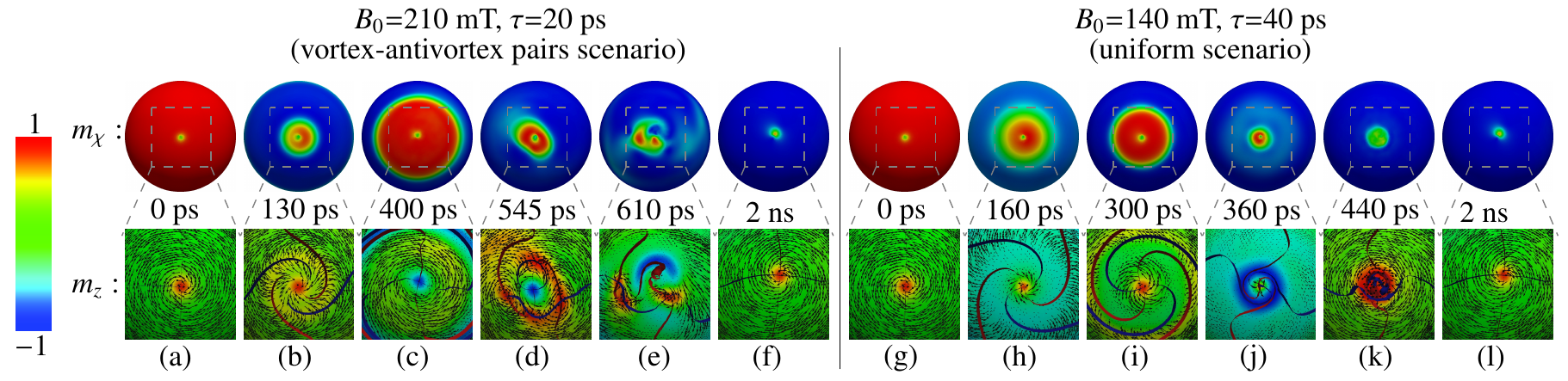}
\caption{(Color online) Mechanisms of vortex chirality switching in a Py hemisphere of inner radius 100~nm and 10~nm thickness. The top row shows the azimuthal component~($m_{\chi}$) of magnetization $\vec m $ in innitial state~(a,g), and dynamics indused by the magnetic pulse~(b-f,h-l). The bottom row  shows the vertical component ($m_z$) of magnetization of the central region of the sample. The blue and red ribbons represent the $m_x=0$ and $m_y=0$ isosurfaces\cite{Hertel07}, respectively; intersection of isosurfaces determine the position of vortex and antivortex cores. Insets (a)-(f) and (g)-(l) correspond to a votrex-antivortex pairs and uniform scenarios, respectively.}\label{fig:types_of_switching}
\end{figure*}

To study the magnetization dynamics induced by the field pulse we perform numerical simulation of the Landau-Lifshitz equation using the \textsc{NMAG} code.\cite{Fischbacher07} We start from the hemispherical shell, see Fig.~\ref{fig:switching-scheme} with inner radius $R_0=100$ nm, radial thickness $h=10$ nm and material parameters of Permalloy ($\mathrm{Ni}_{80}\mathrm{Fe}_{20}$), namely the saturation magnetization $M_s=7.96\times10^5$~A/m, exchange constant $A=1.3\times10^{-11}$~J/m, anisotropy is neglected, and damping coefficient $\alpha=0.01$. The chosen material parameters result in characteristic length scale of the system, an exchange length, $\ell=5.7$ nm. The volume domain of the sample is discretized using irregular tetrahedral mesh with cell size about 3.5~nm.

At the first step, we obtain the equilibrium vortex magnetization distribution by applying the numerical procedure of the total energy minimization. In accordance with the recently obtained phase diagram of equilibrium states of soft magnetic hemispherical shells\cite{Sheka13b} the vortex distribution is the ground state of the system with the referred above size parameters.

Time evolution of the magnetization subsystem disturbed by the applied field pulse can be quite different depending on the pulse parameters. We classify possible types of the magnetization behavior using the time dependent integral chirality
$$  C(t)=\frac{1}{V}\int m_\chi(\vec r,t) \mathrm{d}\vec r, $$
where $m_\chi$ is the azimuthal magnetization component in the spherical frame of reference $\vec m=(m_\vartheta,\,m_\chi,\,m_r)$ and $V$ is volume of the shell. For the equilibrium vortex states $C=\pm1$, see Fig.~\ref{fig:switching-scheme}(a,b). By varying the pulse parameters $B_0$ and $\tau$ in wide ranges we obtain the switching diagram presented in the Fig.~\ref{fig:diagrams}(a). Two domains of vortex chirality switching can be distinguished, namely controllable area where the chirality $C(t)$ changes the sign only once (filled boxes), and potentially uncontrollable area where the chirality $C(t)$ changes sign odd number of times in switching (filled triangles). As it follows from Fig.~\ref{fig:diagrams}(a), the optimal pulse widths for the chirality switching is 30-70 ps which coincides with the pulse width needed for the vortex polarity switching\cite{Hertel07} however the optimal pulse amplitude is larger than for the case of the polarity switching\cite{Hertel07}.

In order to clarify the curvature effect in the chirality switching process we consider a set of truncated spherical caps with the constant volume $V$ and thickness $h$, see Fig.~3 in the Ref.~\onlinecite{Sloika14}. Thereby, the curvature radius $R=R_0/\sqrt{1-\cos\vartheta_c}$ is connected with the polar truncation angle $\vartheta_c$ and it can be varied in the range $R_0\le R<\infty$ with the corresponding variation of the truncation angle $\pi/2\ge\vartheta_c>0$. The singular case $\vartheta_c=0$ corresponds to the planar disk with curvature radius $R\to\infty$ and straight radius $R_\mathrm{disk}=\sqrt2 R_0$. Here we assume that $h\ll R_0$. Evolution of the switching diagram with the curvature radius $R$ increasing is demonstrated in Fig.~\ref{fig:diagrams}(f,g). We conclude that the pulse width optimal for the switching remains approximately the same while the pulse amplitude increases drastically with the curvature radius increasing. As a result, for the case of planar disk the chirality switching requires high field amplitude with high-accurate control of the pulse with. Thereby we confirm the previous result that the vortex chirality can hardly be controlled in a disk by vertical magnetic pulses.\cite{Hoellinger03}

To clarify the reasons of the established curvature effect on the chirality switching we consider absorption of the energy which the sample gains from the applied field pulse. Three contributions to the total energy $E=E_{ex}+E_{ms}+E_{z}$ are considered, namely exchange $E_{ex}$, magnetostatic $E_{ms}$ and Zeeman $E_z$ interactions. Typical time evolution of the referred energy constituents is presented in the Fig.~\ref{fig:energies}(a). The negative contribution represents the interaction with the applied field and its time profile approximately replicates the field pulse profile. The pulse action results in well pronounced jump of the total energy, see Fig.~\ref{fig:energies}(a), and therefore the amount of the absorbed energy can be estimated as $\Delta E=E_{max}-E_{in}$, where $E_{max}$ and $E_{in}$ denote the maximum and initial values of the total energy, respectively, see Fig.~\ref{fig:energies}(a). The resulting dependence $\Delta E(R)$ shows that decreasing of the curvature radius leads to more effective energy absorbtion. As a result, for a hemispherical shell one needs lower amplitude pulse to overcome the energy barrier separating states with opposite chiralities as compared to the case of planar disks. This increases the controllability of the chirality switching process for the hemispherical shells.

At the next step, the details of the chirality switching mechanism are explored. Initially one has an unperturbed vortex state, see Fig.~\ref{fig:types_of_switching}(a,g). The application of the field pulse can induce the switching process, which consists of three stages. At the the first stage a closed circular 180$^\circ$ domain wall is induced,\footnote{Typically a pair of circular domain walls is nucleated. One of these walls moves to the bottom edge of the cap and annihilates there and another wall converges to the cap pole.} see Fig.~\ref{fig:types_of_switching}(b,h). At the second stage, the induced circular domain wall converges to pole of the hemispherical cap and oscillates in the pole neighborhood, see Fig.~\ref{fig:types_of_switching}(b-c, h-j). At the last stage, the collapse of the circular domain wall, see Fig.~\ref{fig:types_of_switching}(d-e,k), results in the vortex state with the chirality opposite to the initial one Fig.~\ref{fig:types_of_switching}(f,l).

Depending on the pulse parameters, we find out two the following scenarios of the collapse of the circular domain wall:

(i) \emph{Vortex-antivortex pairs scenario} -- in neighborhood of the cap pole the circular domain wall experiences an instability, resulting in breakup of the wall into vortex-antivortex pairs, see, Fig.~\ref{fig:types_of_switching}(d). The similar vortex-antivortex rings (circularly closed cross-tie domain walls) were recently obtained in planar disks under action of the perpendicular spin-current.\cite{Volkov11} The number of vortex-antivortex pairs varies from 1 to 4, depending on the pulse parameters. Then one of the newly born antivortices annihilates with the central vortex, see Fig.~\ref{fig:types_of_switching}(e), and after annihilation of rest of the vortex-antivortex pairs a single vortex remains, its chirality is opposite to chirality of the initial vortex. This mechanism is analogous to the chirality switching of vortex domain wall on a nanotube by means of the vortex-anivortex pair formation\cite{Yan12}.

(ii) \emph{Uniform scenario} -- radius of the circular domain wall is shrunk to the value comparable with the core size of the central vortex. As a result the vortex with large core radius is formed, see Fig.~\ref{fig:types_of_switching}(k). After a short time ($\sim10$ ps) radius of the vortex core comes back to its equilibrium value, however the chirality is changed, see Fig.~\ref{fig:types_of_switching}(l). This mechanism is analogous to the uniform chirality switching of vortex domain wall on a nanotube.\cite{Otalora12a}

The vortex-antivortex pairs scenario of the circular domain wall collapse is typical for border of the controllable switching region and for regions of potentially uncontrollable switching, whereas the uniform scenario is common for inner part of the controllable switching region, see Fig.~\ref{fig:diagrams}(a). It should be also noted that the described chirality switching mechanisms do not allow one to control polarity of the vortex.

In conclusion, using micromagnetic simulations we demonstrate that chirality of the vortex state hemispherical shell can be switched by simple field pulse in controllable way, in contrast to the case of planar disk. Key points of the chirality switching process are nucleation and posterior collapse of the circularly closed domain wall. Collapse of the circular domain wall can be attended by creation of the chain of vortex-antivortex pairs.

Authors acknowledge Dr. Denys Makarov, IFW Dresden for fruitful discussions.


\begin{thebibliography}{31}%
\makeatletter
\providecommand \@ifxundefined [1]{%
 \@ifx{#1\undefined}
}%
\providecommand \@ifnum [1]{%
 \ifnum #1\expandafter \@firstoftwo
 \else \expandafter \@secondoftwo
 \fi
}%
\providecommand \@ifx [1]{%
 \ifx #1\expandafter \@firstoftwo
 \else \expandafter \@secondoftwo
 \fi
}%
\providecommand \natexlab [1]{#1}%
\providecommand \enquote  [1]{``#1''}%
\providecommand \bibnamefont  [1]{#1}%
\providecommand \bibfnamefont [1]{#1}%
\providecommand \citenamefont [1]{#1}%
\providecommand \href@noop [0]{\@secondoftwo}%
\providecommand \href [0]{\begingroup \@sanitize@url \@href}%
\providecommand \@href[1]{\@@startlink{#1}\@@href}%
\providecommand \@@href[1]{\endgroup#1\@@endlink}%
\providecommand \@sanitize@url [0]{\catcode `\\12\catcode `\$12\catcode
  `\&12\catcode `\#12\catcode `\^12\catcode `\_12\catcode `\%12\relax}%
\providecommand \@@startlink[1]{}%
\providecommand \@@endlink[0]{}%
\providecommand \url  [0]{\begingroup\@sanitize@url \@url }%
\providecommand \@url [1]{\endgroup\@href {#1}{\urlprefix }}%
\providecommand \urlprefix  [0]{URL }%
\providecommand \Eprint [0]{\href }%
\providecommand \doibase [0]{http://dx.doi.org/}%
\providecommand \selectlanguage [0]{\@gobble}%
\providecommand \bibinfo  [0]{\@secondoftwo}%
\providecommand \bibfield  [0]{\@secondoftwo}%
\providecommand \translation [1]{[#1]}%
\providecommand \BibitemOpen [0]{}%
\providecommand \bibitemStop [0]{}%
\providecommand \bibitemNoStop [0]{.\EOS\space}%
\providecommand \EOS [0]{\spacefactor3000\relax}%
\providecommand \BibitemShut  [1]{\csname bibitem#1\endcsname}%
\let\auto@bib@innerbib\@empty
\bibitem [{\citenamefont {Hubert}\ and\ \citenamefont {Sch{\"
  a}fer}(1998)}]{Hubert98}%
  \BibitemOpen
  \bibfield  {author} {\bibinfo {author} {\bibfnamefont {A.}~\bibnamefont
  {Hubert}}\ and\ \bibinfo {author} {\bibfnamefont {R.}~\bibnamefont {Sch{\"
  a}fer}},\ }\href@noop {} {\emph {\bibinfo {title} {Magnetic domains: the
  analysis of magnetic microstructures}}}\ (\bibinfo  {publisher}
  {Springer--Verlag},\ \bibinfo {address} {Berlin},\ \bibinfo {year}
  {1998})\BibitemShut {NoStop}%
\bibitem [{\citenamefont {Feldtkeller}\ and\ \citenamefont
  {Thomas}(1965)}]{Feldtkeller65}%
  \BibitemOpen
  \bibfield  {author} {\bibinfo {author} {\bibfnamefont {E.}~\bibnamefont
  {Feldtkeller}}\ and\ \bibinfo {author} {\bibfnamefont {H.}~\bibnamefont
  {Thomas}},\ }\href {http://dx.doi.org/10.1007/BF02423256} {\bibfield
  {journal} {\bibinfo  {journal} {Zeitschrift f{\"u}r Physik B Condensed
  Matter}\ }\textbf {\bibinfo {volume} {4}},\ \bibinfo {pages} {8} (\bibinfo
  {year} {1965})}\BibitemShut {NoStop}%
\bibitem [{\citenamefont {Sheka}\ \emph {et~al.}(2013)\citenamefont {Sheka},
  \citenamefont {Kravchuk}, \citenamefont {Sloika},\ and\ \citenamefont
  {Gaididei}}]{Sheka13b}%
  \BibitemOpen
  \bibfield  {author} {\bibinfo {author} {\bibfnamefont {D.~D.}\ \bibnamefont
  {Sheka}}, \bibinfo {author} {\bibfnamefont {V.~P.}\ \bibnamefont {Kravchuk}},
  \bibinfo {author} {\bibfnamefont {M.~I.}\ \bibnamefont {Sloika}}, \ and\
  \bibinfo {author} {\bibfnamefont {Y.}~\bibnamefont {Gaididei}},\ }\href
  {\doibase 10.1142/S2010324713400031} {\bibfield  {journal} {\bibinfo
  {journal} {SPIN}\ }\textbf {\bibinfo {volume} {3}},\ \bibinfo {pages}
  {1340003} (\bibinfo {year} {2013})}\BibitemShut {NoStop}%
\bibitem [{\citenamefont {Soares}\ \emph {et~al.}(2008)\citenamefont {Soares},
  \citenamefont {de~Biasi}, \citenamefont {Coelho}, \citenamefont {dos Santos},
  \citenamefont {de~Menezes}, \citenamefont {Knobel}, \citenamefont {Sampaio},\
  and\ \citenamefont {Garcia}}]{Soares08}%
  \BibitemOpen
  \bibfield  {author} {\bibinfo {author} {\bibfnamefont {M.~M.}\ \bibnamefont
  {Soares}}, \bibinfo {author} {\bibfnamefont {E.}~\bibnamefont {de~Biasi}},
  \bibinfo {author} {\bibfnamefont {L.~N.}\ \bibnamefont {Coelho}}, \bibinfo
  {author} {\bibfnamefont {M.~C.}\ \bibnamefont {dos Santos}}, \bibinfo
  {author} {\bibfnamefont {F.~S.}\ \bibnamefont {de~Menezes}}, \bibinfo
  {author} {\bibfnamefont {M.}~\bibnamefont {Knobel}}, \bibinfo {author}
  {\bibfnamefont {L.~C.}\ \bibnamefont {Sampaio}}, \ and\ \bibinfo {author}
  {\bibfnamefont {F.}~\bibnamefont {Garcia}},\ }\href {\doibase
  10.1103/PhysRevB.77.224405} {\bibfield  {journal} {\bibinfo  {journal} {Phys.
  Rev. B}\ }\textbf {\bibinfo {volume} {77}},\ \bibinfo {eid} {224405}
  (\bibinfo {year} {2008})}\BibitemShut {NoStop}%
\bibitem [{\citenamefont {Streubel}\ \emph
  {et~al.}(2012{\natexlab{a}})\citenamefont {Streubel}, \citenamefont
  {Kravchuk}, \citenamefont {Sheka}, \citenamefont {Makarov}, \citenamefont
  {Kronast}, \citenamefont {Schmidt},\ and\ \citenamefont
  {Gaididei}}]{Streubel12}%
  \BibitemOpen
  \bibfield  {author} {\bibinfo {author} {\bibfnamefont {R.}~\bibnamefont
  {Streubel}}, \bibinfo {author} {\bibfnamefont {V.~P.}\ \bibnamefont
  {Kravchuk}}, \bibinfo {author} {\bibfnamefont {D.~D.}\ \bibnamefont {Sheka}},
  \bibinfo {author} {\bibfnamefont {D.}~\bibnamefont {Makarov}}, \bibinfo
  {author} {\bibfnamefont {F.}~\bibnamefont {Kronast}}, \bibinfo {author}
  {\bibfnamefont {O.~G.}\ \bibnamefont {Schmidt}}, \ and\ \bibinfo {author}
  {\bibfnamefont {Y.}~\bibnamefont {Gaididei}},\ }\href {\doibase
  10.1063/1.4756708} {\bibfield  {journal} {\bibinfo  {journal} {Appl. Phys.
  Lett.}\ }\textbf {\bibinfo {volume} {101}},\ \bibinfo {eid} {132419}
  (\bibinfo {year} {2012}{\natexlab{a}})}\BibitemShut {NoStop}%
\bibitem [{\citenamefont {Streubel}\ \emph
  {et~al.}(2012{\natexlab{b}})\citenamefont {Streubel}, \citenamefont
  {Makarov}, \citenamefont {Kronast}, \citenamefont {Kravchuk}, \citenamefont
  {Albrecht},\ and\ \citenamefont {Schmidt}}]{Streubel12a}%
  \BibitemOpen
  \bibfield  {author} {\bibinfo {author} {\bibfnamefont {R.}~\bibnamefont
  {Streubel}}, \bibinfo {author} {\bibfnamefont {D.}~\bibnamefont {Makarov}},
  \bibinfo {author} {\bibfnamefont {F.}~\bibnamefont {Kronast}}, \bibinfo
  {author} {\bibfnamefont {V.}~\bibnamefont {Kravchuk}}, \bibinfo {author}
  {\bibfnamefont {M.}~\bibnamefont {Albrecht}}, \ and\ \bibinfo {author}
  {\bibfnamefont {O.~G.}\ \bibnamefont {Schmidt}},\ }\href {\doibase
  10.1103/PhysRevB.85.174429} {\bibfield  {journal} {\bibinfo  {journal} {Phys.
  Rev. B}\ }\textbf {\bibinfo {volume} {85}},\ \bibinfo {pages} {174429}
  (\bibinfo {year} {2012}{\natexlab{b}})}\BibitemShut {NoStop}%
\bibitem [{\citenamefont {Bohlens}\ \emph {et~al.}(2008)\citenamefont
  {Bohlens}, \citenamefont {Kr\"{u}ger}, \citenamefont {Drews}, \citenamefont
  {Bolte}, \citenamefont {Meier},\ and\ \citenamefont
  {Pfannkuche}}]{Bohlens08}%
  \BibitemOpen
  \bibfield  {author} {\bibinfo {author} {\bibfnamefont {S.}~\bibnamefont
  {Bohlens}}, \bibinfo {author} {\bibfnamefont {B.}~\bibnamefont {Kr\"{u}ger}},
  \bibinfo {author} {\bibfnamefont {A.}~\bibnamefont {Drews}}, \bibinfo
  {author} {\bibfnamefont {M.}~\bibnamefont {Bolte}}, \bibinfo {author}
  {\bibfnamefont {G.}~\bibnamefont {Meier}}, \ and\ \bibinfo {author}
  {\bibfnamefont {D.}~\bibnamefont {Pfannkuche}},\ }\href {\doibase
  10.1063/1.2998584} {\bibfield  {journal} {\bibinfo  {journal} {Appl. Phys.
  Lett.}\ }\textbf {\bibinfo {volume} {93}},\ \bibinfo {eid} {142508} (\bibinfo
  {year} {2008})}\BibitemShut {NoStop}%
\bibitem [{\citenamefont {Pigeau}\ \emph {et~al.}(2010)\citenamefont {Pigeau},
  \citenamefont {de~Loubens}, \citenamefont {Klein}, \citenamefont {Riegler},
  \citenamefont {Lochner}, \citenamefont {Schmidt}, \citenamefont {Molenkamp},
  \citenamefont {Tiberkevich},\ and\ \citenamefont {Slavin}}]{Pigeau10}%
  \BibitemOpen
  \bibfield  {author} {\bibinfo {author} {\bibfnamefont {B.}~\bibnamefont
  {Pigeau}}, \bibinfo {author} {\bibfnamefont {G.}~\bibnamefont {de~Loubens}},
  \bibinfo {author} {\bibfnamefont {O.}~\bibnamefont {Klein}}, \bibinfo
  {author} {\bibfnamefont {A.}~\bibnamefont {Riegler}}, \bibinfo {author}
  {\bibfnamefont {F.}~\bibnamefont {Lochner}}, \bibinfo {author} {\bibfnamefont
  {G.}~\bibnamefont {Schmidt}}, \bibinfo {author} {\bibfnamefont {L.~W.}\
  \bibnamefont {Molenkamp}}, \bibinfo {author} {\bibfnamefont {V.~S.}\
  \bibnamefont {Tiberkevich}}, \ and\ \bibinfo {author} {\bibfnamefont {A.~N.}\
  \bibnamefont {Slavin}},\ }\href {\doibase DOI:10.1063/1.3373833} {\bibfield
  {journal} {\bibinfo  {journal} {Appl. Phys. Lett.}\ }\textbf {\bibinfo
  {volume} {96}},\ \bibinfo {pages} {132506} (\bibinfo {year}
  {2010})}\BibitemShut {NoStop}%
\bibitem [{\citenamefont {Yu}\ \emph {et~al.}(2011)\citenamefont {Yu},
  \citenamefont {Jung}, \citenamefont {Lee}, \citenamefont {Fischer},\ and\
  \citenamefont {Kim}}]{Yu11a}%
  \BibitemOpen
  \bibfield  {author} {\bibinfo {author} {\bibfnamefont {Y.-S.}\ \bibnamefont
  {Yu}}, \bibinfo {author} {\bibfnamefont {H.}~\bibnamefont {Jung}}, \bibinfo
  {author} {\bibfnamefont {K.-S.}\ \bibnamefont {Lee}}, \bibinfo {author}
  {\bibfnamefont {P.}~\bibnamefont {Fischer}}, \ and\ \bibinfo {author}
  {\bibfnamefont {S.-K.}\ \bibnamefont {Kim}},\ }\href {\doibase
  10.1063/1.3551524} {\bibfield  {journal} {\bibinfo  {journal} {Appl. Phys.
  Lett.}\ }\textbf {\bibinfo {volume} {98}},\ \bibinfo {eid} {052507} (\bibinfo
  {year} {2011})}\BibitemShut {NoStop}%
\bibitem [{\citenamefont {Nakano}\ \emph {et~al.}(2011)\citenamefont {Nakano},
  \citenamefont {Chiba}, \citenamefont {Ohshima}, \citenamefont {Kasai},
  \citenamefont {Sato}, \citenamefont {Nakatani}, \citenamefont {Sekiguchi},
  \citenamefont {Kobayashi},\ and\ \citenamefont {Ono}}]{Nakano11}%
  \BibitemOpen
  \bibfield  {author} {\bibinfo {author} {\bibfnamefont {K.}~\bibnamefont
  {Nakano}}, \bibinfo {author} {\bibfnamefont {D.}~\bibnamefont {Chiba}},
  \bibinfo {author} {\bibfnamefont {N.}~\bibnamefont {Ohshima}}, \bibinfo
  {author} {\bibfnamefont {S.}~\bibnamefont {Kasai}}, \bibinfo {author}
  {\bibfnamefont {T.}~\bibnamefont {Sato}}, \bibinfo {author} {\bibfnamefont
  {Y.}~\bibnamefont {Nakatani}}, \bibinfo {author} {\bibfnamefont
  {K.}~\bibnamefont {Sekiguchi}}, \bibinfo {author} {\bibfnamefont
  {K.}~\bibnamefont {Kobayashi}}, \ and\ \bibinfo {author} {\bibfnamefont
  {T.}~\bibnamefont {Ono}},\ }\href {\doibase 10.1063/1.3673303} {\bibfield
  {journal} {\bibinfo  {journal} {Appl. Phys. Lett.}\ }\textbf {\bibinfo
  {volume} {99}},\ \bibinfo {eid} {262505} (\bibinfo {year}
  {2011})}\BibitemShut {NoStop}%
\bibitem [{\citenamefont {Van~Waeyenberge}\ \emph {et~al.}(2006)\citenamefont
  {Van~Waeyenberge}, \citenamefont {Puzic}, \citenamefont {Stoll},
  \citenamefont {Chou}, \citenamefont {Tyliszczak}, \citenamefont {Hertel},
  \citenamefont {F\"ahnle}, \citenamefont {Bruckl}, \citenamefont {Rott},
  \citenamefont {Reiss}, \citenamefont {Neudecker}, \citenamefont {Weiss},
  \citenamefont {Back},\ and\ \citenamefont {Sch\"utz}}]{Waeyenberge06}%
  \BibitemOpen
  \bibfield  {author} {\bibinfo {author} {\bibfnamefont {B.}~\bibnamefont
  {Van~Waeyenberge}}, \bibinfo {author} {\bibfnamefont {A.}~\bibnamefont
  {Puzic}}, \bibinfo {author} {\bibfnamefont {H.}~\bibnamefont {Stoll}},
  \bibinfo {author} {\bibfnamefont {K.~W.}\ \bibnamefont {Chou}}, \bibinfo
  {author} {\bibfnamefont {T.}~\bibnamefont {Tyliszczak}}, \bibinfo {author}
  {\bibfnamefont {R.}~\bibnamefont {Hertel}}, \bibinfo {author} {\bibfnamefont
  {M.}~\bibnamefont {F\"ahnle}}, \bibinfo {author} {\bibfnamefont
  {H.}~\bibnamefont {Bruckl}}, \bibinfo {author} {\bibfnamefont
  {K.}~\bibnamefont {Rott}}, \bibinfo {author} {\bibfnamefont {G.}~\bibnamefont
  {Reiss}}, \bibinfo {author} {\bibfnamefont {I.}~\bibnamefont {Neudecker}},
  \bibinfo {author} {\bibfnamefont {D.}~\bibnamefont {Weiss}}, \bibinfo
  {author} {\bibfnamefont {C.~H.}\ \bibnamefont {Back}}, \ and\ \bibinfo
  {author} {\bibfnamefont {G.}~\bibnamefont {Sch\"utz}},\ }\href
  {http://dx.doi.org/10.1038/nature05240} {\bibfield  {journal} {\bibinfo
  {journal} {Nature}\ }\textbf {\bibinfo {volume} {444}},\ \bibinfo {pages}
  {461} (\bibinfo {year} {2006})}\BibitemShut {NoStop}%
\bibitem [{\citenamefont {Hertel}\ \emph {et~al.}(2007)\citenamefont {Hertel},
  \citenamefont {Gliga}, \citenamefont {F\"ahnle},\ and\ \citenamefont
  {Schneider}}]{Hertel07}%
  \BibitemOpen
  \bibfield  {author} {\bibinfo {author} {\bibfnamefont {R.}~\bibnamefont
  {Hertel}}, \bibinfo {author} {\bibfnamefont {S.}~\bibnamefont {Gliga}},
  \bibinfo {author} {\bibfnamefont {M.}~\bibnamefont {F\"ahnle}}, \ and\
  \bibinfo {author} {\bibfnamefont {C.~M.}\ \bibnamefont {Schneider}},\ }\href
  {http://link.aps.org/abstract/PRL/v98/e117201} {\bibfield  {journal}
  {\bibinfo  {journal} {Phys. Rev. Lett.}\ }\textbf {\bibinfo {volume} {98}},\
  \bibinfo {eid} {117201} (\bibinfo {year} {2007})}\BibitemShut {NoStop}%
\bibitem [{\citenamefont {Kravchuk}\ \emph {et~al.}(2007)\citenamefont
  {Kravchuk}, \citenamefont {Sheka}, \citenamefont {Gaididei},\ and\
  \citenamefont {Mertens}}]{Kravchuk07c}%
  \BibitemOpen
  \bibfield  {author} {\bibinfo {author} {\bibfnamefont {V.~P.}\ \bibnamefont
  {Kravchuk}}, \bibinfo {author} {\bibfnamefont {D.~D.}\ \bibnamefont {Sheka}},
  \bibinfo {author} {\bibfnamefont {Y.}~\bibnamefont {Gaididei}}, \ and\
  \bibinfo {author} {\bibfnamefont {F.~G.}\ \bibnamefont {Mertens}},\ }\href
  {\doibase 10.1063/1.2770819} {\bibfield  {journal} {\bibinfo  {journal}
  {J.~Appl. Phys.}\ }\textbf {\bibinfo {volume} {102}},\ \bibinfo {eid}
  {043908} (\bibinfo {year} {2007})}\BibitemShut {NoStop}%
\bibitem [{\citenamefont {Curcic}\ \emph {et~al.}(2008)\citenamefont {Curcic},
  \citenamefont {Van~Waeyenberge}, \citenamefont {Vansteenkiste}, \citenamefont
  {Weigand}, \citenamefont {Sackmann}, \citenamefont {Stoll}, \citenamefont
  {F\"{a}hnle}, \citenamefont {Tyliszczak}, \citenamefont {Woltersdorf},
  \citenamefont {Back},\ and\ \citenamefont {Sch\"{u}tz}}]{Curcic08a}%
  \BibitemOpen
  \bibfield  {author} {\bibinfo {author} {\bibfnamefont {M.}~\bibnamefont
  {Curcic}}, \bibinfo {author} {\bibfnamefont {B.}~\bibnamefont
  {Van~Waeyenberge}}, \bibinfo {author} {\bibfnamefont {A.}~\bibnamefont
  {Vansteenkiste}}, \bibinfo {author} {\bibfnamefont {M.}~\bibnamefont
  {Weigand}}, \bibinfo {author} {\bibfnamefont {V.}~\bibnamefont {Sackmann}},
  \bibinfo {author} {\bibfnamefont {H.}~\bibnamefont {Stoll}}, \bibinfo
  {author} {\bibfnamefont {M.}~\bibnamefont {F\"{a}hnle}}, \bibinfo {author}
  {\bibfnamefont {T.}~\bibnamefont {Tyliszczak}}, \bibinfo {author}
  {\bibfnamefont {G.}~\bibnamefont {Woltersdorf}}, \bibinfo {author}
  {\bibfnamefont {C.~H.}\ \bibnamefont {Back}}, \ and\ \bibinfo {author}
  {\bibfnamefont {G.}~\bibnamefont {Sch\"{u}tz}},\ }\href {\doibase
  10.1103/PhysRevLett.101.197204} {\bibfield  {journal} {\bibinfo  {journal}
  {Phys. Rev. Lett.}\ }\textbf {\bibinfo {volume} {101}},\ \bibinfo {eid}
  {197204} (\bibinfo {year} {2008})}\BibitemShut {NoStop}%
\bibitem [{\citenamefont {Kammerer}\ \emph {et~al.}(2011)\citenamefont
  {Kammerer}, \citenamefont {Weigand}, \citenamefont {Curcic}, \citenamefont
  {Noske}, \citenamefont {Sproll}, \citenamefont {Vansteenkiste}, \citenamefont
  {Van~Waeyenberge}, \citenamefont {Stoll}, \citenamefont {Woltersdorf},
  \citenamefont {Back},\ and\ \citenamefont {Schuetz}}]{Kammerer11}%
  \BibitemOpen
  \bibfield  {author} {\bibinfo {author} {\bibfnamefont {M.}~\bibnamefont
  {Kammerer}}, \bibinfo {author} {\bibfnamefont {M.}~\bibnamefont {Weigand}},
  \bibinfo {author} {\bibfnamefont {M.}~\bibnamefont {Curcic}}, \bibinfo
  {author} {\bibfnamefont {M.}~\bibnamefont {Noske}}, \bibinfo {author}
  {\bibfnamefont {M.}~\bibnamefont {Sproll}}, \bibinfo {author} {\bibfnamefont
  {A.}~\bibnamefont {Vansteenkiste}}, \bibinfo {author} {\bibfnamefont
  {B.}~\bibnamefont {Van~Waeyenberge}}, \bibinfo {author} {\bibfnamefont
  {H.}~\bibnamefont {Stoll}}, \bibinfo {author} {\bibfnamefont
  {G.}~\bibnamefont {Woltersdorf}}, \bibinfo {author} {\bibfnamefont {C.~H.}\
  \bibnamefont {Back}}, \ and\ \bibinfo {author} {\bibfnamefont
  {G.}~\bibnamefont {Schuetz}},\ }\href {http://dx.doi.org/10.1038/ncomms1277}
  {\bibfield  {journal} {\bibinfo  {journal} {Nat Commun}\ }\textbf {\bibinfo
  {volume} {2}},\ \bibinfo {pages} {279} (\bibinfo {year} {2011})}\BibitemShut
  {NoStop}%
\bibitem [{\citenamefont {Caputo}\ \emph {et~al.}(2007)\citenamefont {Caputo},
  \citenamefont {Gaididei}, \citenamefont {Mertens},\ and\ \citenamefont
  {Sheka}}]{Caputo07}%
  \BibitemOpen
  \bibfield  {author} {\bibinfo {author} {\bibfnamefont {J.-G.}\ \bibnamefont
  {Caputo}}, \bibinfo {author} {\bibfnamefont {Y.}~\bibnamefont {Gaididei}},
  \bibinfo {author} {\bibfnamefont {F.~G.}\ \bibnamefont {Mertens}}, \ and\
  \bibinfo {author} {\bibfnamefont {D.~D.}\ \bibnamefont {Sheka}},\ }\href
  {http://link.aps.org/abstract/PRL/v98/e056604} {\bibfield  {journal}
  {\bibinfo  {journal} {Phys. Rev. Lett.}\ }\textbf {\bibinfo {volume} {98}},\
  \bibinfo {eid} {056604} (\bibinfo {year} {2007})}\BibitemShut {NoStop}%
\bibitem [{\citenamefont {Yamada}\ \emph {et~al.}(2007)\citenamefont {Yamada},
  \citenamefont {Kasai}, \citenamefont {Nakatani}, \citenamefont {Kobayashi},
  \citenamefont {Kohno}, \citenamefont {Thiaville},\ and\ \citenamefont
  {Ono}}]{Yamada07}%
  \BibitemOpen
  \bibfield  {author} {\bibinfo {author} {\bibfnamefont {K.}~\bibnamefont
  {Yamada}}, \bibinfo {author} {\bibfnamefont {S.}~\bibnamefont {Kasai}},
  \bibinfo {author} {\bibfnamefont {Y.}~\bibnamefont {Nakatani}}, \bibinfo
  {author} {\bibfnamefont {K.}~\bibnamefont {Kobayashi}}, \bibinfo {author}
  {\bibfnamefont {H.}~\bibnamefont {Kohno}}, \bibinfo {author} {\bibfnamefont
  {A.}~\bibnamefont {Thiaville}}, \ and\ \bibinfo {author} {\bibfnamefont
  {T.}~\bibnamefont {Ono}},\ }\href {http://dx.doi.org/10.1038/nmat1867}
  {\bibfield  {journal} {\bibinfo  {journal} {Nat Mater}\ }\textbf {\bibinfo
  {volume} {6}},\ \bibinfo {pages} {270} (\bibinfo {year} {2007})}\BibitemShut
  {NoStop}%
\bibitem [{\citenamefont {Jaafar}\ \emph {et~al.}(2010)\citenamefont {Jaafar},
  \citenamefont {Yanes}, \citenamefont {Perez~de Lara}, \citenamefont
  {Chubykalo-Fesenko}, \citenamefont {Asenjo}, \citenamefont {Gonzalez},
  \citenamefont {Anguita}, \citenamefont {Vazquez},\ and\ \citenamefont
  {Vicent}}]{Jaafar10}%
  \BibitemOpen
  \bibfield  {author} {\bibinfo {author} {\bibfnamefont {M.}~\bibnamefont
  {Jaafar}}, \bibinfo {author} {\bibfnamefont {R.}~\bibnamefont {Yanes}},
  \bibinfo {author} {\bibfnamefont {D.}~\bibnamefont {Perez~de Lara}}, \bibinfo
  {author} {\bibfnamefont {O.}~\bibnamefont {Chubykalo-Fesenko}}, \bibinfo
  {author} {\bibfnamefont {A.}~\bibnamefont {Asenjo}}, \bibinfo {author}
  {\bibfnamefont {E.~M.}\ \bibnamefont {Gonzalez}}, \bibinfo {author}
  {\bibfnamefont {J.~V.}\ \bibnamefont {Anguita}}, \bibinfo {author}
  {\bibfnamefont {M.}~\bibnamefont {Vazquez}}, \ and\ \bibinfo {author}
  {\bibfnamefont {J.~L.}\ \bibnamefont {Vicent}},\ }\href {\doibase
  10.1103/PhysRevB.81.054439} {\bibfield  {journal} {\bibinfo  {journal} {Phys.
  Rev. B}\ }\textbf {\bibinfo {volume} {81}},\ \bibinfo {pages} {054439}
  (\bibinfo {year} {2010})}\BibitemShut {NoStop}%
\bibitem [{\citenamefont {Uhl\'{\i}\u{r}}\ \emph {et~al.}(2013)\citenamefont
  {Uhl\'{\i}\u{r}}, \citenamefont {Urb\'{a}nek}, \citenamefont {Hlad\'{\i}k},
  \citenamefont {Spousta}, \citenamefont {Im}, \citenamefont {Fischer},
  \citenamefont {Eibagi}, \citenamefont {Kan},\ and\ \citenamefont
  {\u{S}ikola}}]{Uhlir13}%
  \BibitemOpen
  \bibfield  {author} {\bibinfo {author} {\bibfnamefont {V.}~\bibnamefont
  {Uhl\'{\i}\u{r}}}, \bibinfo {author} {\bibfnamefont {M.}~\bibnamefont
  {Urb\'{a}nek}}, \bibinfo {author} {\bibfnamefont {L.}~\bibnamefont
  {Hlad\'{\i}k}}, \bibinfo {author} {\bibfnamefont {J.}~\bibnamefont
  {Spousta}}, \bibinfo {author} {\bibfnamefont {M.-Y.}\ \bibnamefont {Im}},
  \bibinfo {author} {\bibfnamefont {P.}~\bibnamefont {Fischer}}, \bibinfo
  {author} {\bibfnamefont {N.}~\bibnamefont {Eibagi}}, \bibinfo {author}
  {\bibfnamefont {E.~E.}\ \bibnamefont {Kan}, \bibfnamefont {J.~J.~Fullerton}},
  \ and\ \bibinfo {author} {\bibfnamefont {T.}~\bibnamefont {\u{S}ikola}},\
  }\href {http://dx.doi.org/10.1038/nnano.2013.66} {\bibfield  {journal}
  {\bibinfo  {journal} {Nat Nano}\ }\textbf {\bibinfo {volume} {8}},\ \bibinfo
  {pages} {341} (\bibinfo {year} {2013})}\BibitemShut {NoStop}%
\bibitem [{\citenamefont {Hertel}(2013)}]{Hertel13}%
  \BibitemOpen
  \bibfield  {author} {\bibinfo {author} {\bibfnamefont {R.}~\bibnamefont
  {Hertel}},\ }\href {http://dx.doi.org/10.1038/nnano.2013.81} {\bibfield
  {journal} {\bibinfo  {journal} {Nat Nano}\ }\textbf {\bibinfo {volume} {8}},\
  \bibinfo {pages} {318} (\bibinfo {year} {2013})}\BibitemShut {NoStop}%
\bibitem [{\citenamefont {Lua}\ \emph {et~al.}(2008)\citenamefont {Lua},
  \citenamefont {Kushvaha}, \citenamefont {Wu}, \citenamefont {Teo},\ and\
  \citenamefont {Chong}}]{Lua08}%
  \BibitemOpen
  \bibfield  {author} {\bibinfo {author} {\bibfnamefont {S.~Y.~H.}\
  \bibnamefont {Lua}}, \bibinfo {author} {\bibfnamefont {S.~S.}\ \bibnamefont
  {Kushvaha}}, \bibinfo {author} {\bibfnamefont {Y.~H.}\ \bibnamefont {Wu}},
  \bibinfo {author} {\bibfnamefont {K.~L.}\ \bibnamefont {Teo}}, \ and\
  \bibinfo {author} {\bibfnamefont {T.~C.}\ \bibnamefont {Chong}},\ }\href
  {\doibase 10.1063/1.2989135} {\bibfield  {journal} {\bibinfo  {journal}
  {Applied Physics Letters}\ }\textbf {\bibinfo {volume} {93}},\ \bibinfo {eid}
  {122504} (\bibinfo {year} {2008})}\BibitemShut {NoStop}%
\bibitem [{\citenamefont {Gaididei}, \citenamefont {Sheka},\ and\ \citenamefont
  {Mertens}(2008)}]{Gaididei08}%
  \BibitemOpen
  \bibfield  {author} {\bibinfo {author} {\bibfnamefont {Y.}~\bibnamefont
  {Gaididei}}, \bibinfo {author} {\bibfnamefont {D.~D.}\ \bibnamefont {Sheka}},
  \ and\ \bibinfo {author} {\bibfnamefont {F.~G.}\ \bibnamefont {Mertens}},\
  }\href {\doibase 10.1063/1.2829795} {\bibfield  {journal} {\bibinfo
  {journal} {Appl. Phys. Lett.}\ }\textbf {\bibinfo {volume} {92}},\ \bibinfo
  {eid} {012503} (\bibinfo {year} {2008})}\BibitemShut {NoStop}%
\bibitem [{\citenamefont {Antos}\ and\ \citenamefont {Otani}(2009)}]{Antos09}%
  \BibitemOpen
  \bibfield  {author} {\bibinfo {author} {\bibfnamefont {R.}~\bibnamefont
  {Antos}}\ and\ \bibinfo {author} {\bibfnamefont {Y.}~\bibnamefont {Otani}},\
  }\href {\doibase 10.1103/PhysRevB.80.140404} {\bibfield  {journal} {\bibinfo
  {journal} {Phys. Rev. B}\ }\textbf {\bibinfo {volume} {80}},\ \bibinfo {eid}
  {140404} (\bibinfo {year} {2009})}\BibitemShut {NoStop}%
\bibitem [{\citenamefont {Choi}\ \emph {et~al.}(2010)\citenamefont {Choi},
  \citenamefont {Yoo}, \citenamefont {Lee}, \citenamefont {Yu}, \citenamefont
  {Jung},\ and\ \citenamefont {Kim}}]{Choi10}%
  \BibitemOpen
  \bibfield  {author} {\bibinfo {author} {\bibfnamefont {Y.-S.}\ \bibnamefont
  {Choi}}, \bibinfo {author} {\bibfnamefont {M.-W.}\ \bibnamefont {Yoo}},
  \bibinfo {author} {\bibfnamefont {K.-S.}\ \bibnamefont {Lee}}, \bibinfo
  {author} {\bibfnamefont {Y.-S.}\ \bibnamefont {Yu}}, \bibinfo {author}
  {\bibfnamefont {H.}~\bibnamefont {Jung}}, \ and\ \bibinfo {author}
  {\bibfnamefont {S.-K.}\ \bibnamefont {Kim}},\ }\href {\doibase
  10.1063/1.3310017} {\bibfield  {journal} {\bibinfo  {journal} {Appl. Phys.
  Lett.}\ }\textbf {\bibinfo {volume} {96}},\ \bibinfo {eid} {072507} (\bibinfo
  {year} {2010})}\BibitemShut {NoStop}%
\bibitem [{\citenamefont {Sloika}\ \emph {et~al.}(2014)\citenamefont {Sloika},
  \citenamefont {Kravchuk}, \citenamefont {Sheka},\ and\ \citenamefont
  {Gaididei}}]{Sloika14}%
  \BibitemOpen
  \bibfield  {author} {\bibinfo {author} {\bibfnamefont {M.~I.}\ \bibnamefont
  {Sloika}}, \bibinfo {author} {\bibfnamefont {V.~P.}\ \bibnamefont
  {Kravchuk}}, \bibinfo {author} {\bibfnamefont {D.~D.}\ \bibnamefont {Sheka}},
  \ and\ \bibinfo {author} {\bibfnamefont {Y.}~\bibnamefont {Gaididei}},\
  }\href {\doibase http://dx.doi.org/10.1063/1.4884957} {\bibfield  {journal}
  {\bibinfo  {journal} {Applied Physics Letters}\ }\textbf {\bibinfo {volume}
  {104}},\ \bibinfo {eid} {252403} (\bibinfo {year} {2014})}\BibitemShut
  {NoStop}%
\bibitem [{\citenamefont {Fischbacher}\ \emph {et~al.}(2007)\citenamefont
  {Fischbacher}, \citenamefont {Franchin}, \citenamefont {Bordignon},\ and\
  \citenamefont {Fangohr}}]{Fischbacher07}%
  \BibitemOpen
  \bibfield  {author} {\bibinfo {author} {\bibfnamefont {T.}~\bibnamefont
  {Fischbacher}}, \bibinfo {author} {\bibfnamefont {M.}~\bibnamefont
  {Franchin}}, \bibinfo {author} {\bibfnamefont {G.}~\bibnamefont {Bordignon}},
  \ and\ \bibinfo {author} {\bibfnamefont {H.}~\bibnamefont {Fangohr}},\ }\href
  {\doibase 10.1109/tmag.2007.893843} {\bibfield  {journal} {\bibinfo
  {journal} {IEEE Trans. Magn.}\ }\textbf {\bibinfo {volume} {43}},\ \bibinfo
  {pages} {2896} (\bibinfo {year} {2007})}\BibitemShut {NoStop}%
\bibitem [{\citenamefont {H{\"o}llinger}, \citenamefont {Killinger},\ and\
  \citenamefont {Krey}(2003)}]{Hoellinger03}%
  \BibitemOpen
  \bibfield  {author} {\bibinfo {author} {\bibfnamefont {R.}~\bibnamefont
  {H{\"o}llinger}}, \bibinfo {author} {\bibfnamefont {A.}~\bibnamefont
  {Killinger}}, \ and\ \bibinfo {author} {\bibfnamefont {U.}~\bibnamefont
  {Krey}},\ }\href
  {http://www.sciencedirect.com/science/article/B6TJJ-47F1D5K-1/2/02a9f675dac9cbb50d643664d50b98be}
  {\bibfield  {journal} {\bibinfo  {journal} {J.~Magn. Magn. Mater.}\ }\textbf
  {\bibinfo {volume} {261}},\ \bibinfo {pages} {178} (\bibinfo {year}
  {2003})}\BibitemShut {NoStop}%
\bibitem [{Note1()}]{Note1}%
  \BibitemOpen
  \bibinfo {note} {Typically a pair of circular domain walls is nucleated. One
  of these walls moves to the bottom edge of the cap and annihilates there and
  another wall converges to the cap pole.}\BibitemShut {Stop}%
\bibitem [{\citenamefont {Volkov}\ \emph {et~al.}(2011)\citenamefont {Volkov},
  \citenamefont {Kravchuk}, \citenamefont {Sheka},\ and\ \citenamefont
  {Gaididei}}]{Volkov11}%
  \BibitemOpen
  \bibfield  {author} {\bibinfo {author} {\bibfnamefont {O.~M.}\ \bibnamefont
  {Volkov}}, \bibinfo {author} {\bibfnamefont {V.~P.}\ \bibnamefont
  {Kravchuk}}, \bibinfo {author} {\bibfnamefont {D.~D.}\ \bibnamefont {Sheka}},
  \ and\ \bibinfo {author} {\bibfnamefont {Y.}~\bibnamefont {Gaididei}},\
  }\href {\doibase 10.1103/PhysRevB.84.052404} {\bibfield  {journal} {\bibinfo
  {journal} {Phys. Rev. B}\ }\textbf {\bibinfo {volume} {84}},\ \bibinfo
  {pages} {052404} (\bibinfo {year} {2011})}\BibitemShut {NoStop}%
\bibitem [{\citenamefont {Yan}\ \emph {et~al.}(2012)\citenamefont {Yan},
  \citenamefont {Andreas}, \citenamefont {Kakay}, \citenamefont
  {Garcia-Sanchez},\ and\ \citenamefont {Hertel}}]{Yan12}%
  \BibitemOpen
  \bibfield  {author} {\bibinfo {author} {\bibfnamefont {M.}~\bibnamefont
  {Yan}}, \bibinfo {author} {\bibfnamefont {C.}~\bibnamefont {Andreas}},
  \bibinfo {author} {\bibfnamefont {A.}~\bibnamefont {Kakay}}, \bibinfo
  {author} {\bibfnamefont {F.}~\bibnamefont {Garcia-Sanchez}}, \ and\ \bibinfo
  {author} {\bibfnamefont {R.}~\bibnamefont {Hertel}},\ }\href {\doibase
  10.1063/1.4727909} {\bibfield  {journal} {\bibinfo  {journal} {Applied
  Physics Letters}\ }\textbf {\bibinfo {volume} {100}},\ \bibinfo {eid}
  {252401} (\bibinfo {year} {2012})}\BibitemShut {NoStop}%
\bibitem [{\citenamefont {Ot\'{a}lora}\ \emph {et~al.}(2012)\citenamefont
  {Ot\'{a}lora}, \citenamefont {J.A.}, \citenamefont {Vargas},\ and\
  \citenamefont {Landeros}}]{Otalora12a}%
  \BibitemOpen
  \bibfield  {author} {\bibinfo {author} {\bibfnamefont {J.}~\bibnamefont
  {Ot\'{a}lora}}, \bibinfo {author} {\bibfnamefont {L.-L.}\ \bibnamefont
  {J.A.}}, \bibinfo {author} {\bibfnamefont {P.}~\bibnamefont {Vargas}}, \ and\
  \bibinfo {author} {\bibfnamefont {P.}~\bibnamefont {Landeros}},\ }\href
  {\doibase http://dx.doi.org/10.1063/1.3687154} {\bibfield  {journal}
  {\bibinfo  {journal} {Applied Physics Letters}\ }\textbf {\bibinfo {volume}
  {100}},\ \bibinfo {eid} {072407} (\bibinfo {year} {2012})}\BibitemShut
  {NoStop}%
\end{thebibliography}

%

\end{document}